# Experimental and Numerical Understanding of Localized Spin Wave Mode Behavior in Broadly Tunable Spatially Complex Magnetic Configurations

Chunhui Du, Rohan Adur, Hailong Wang, Sergei A. Manuilov, Fengyuan Yang,

Denis V. Pelekhov*, and P. Chris Hammel*

Department of Physics, The Ohio State University, Columbus, OH, 43210, USA

*Emails: pelekhov.1@osu.edu; hammel@physics.osu.edu

Abstract

Spin wave modes confined in a ferromagnetic film by the spatially inhomogeneous magnetic field generated by a scanned micromagnetic tip of a ferromagnetic resonance force microscope (FMRFM) enable microscopic imaging of the internal fields and spin dynamics in nanoscale magnetic devices. Here we report a detailed study of spin wave modes in a thin ferromagnetic film localized by magnetic field configurations frequently encountered in FMRFM experiments, including geometries in which the probe magnetic moment is both parallel and antiparallel to the applied uniform magnetic field. We demonstrate that characteristics of the localized modes, such as resonance field and confinement radius, can be broadly tuned by controlling the orientation of the applied field relative to the film plane. Micromagnetic simulations accurately reproduce our FMRFM spectra allowing quantitative understanding of the localized modes. Our results reveal a general method of generating tightly confined spin wave modes in various geometries with excellent spatial resolution that significantly facilitates the broad application of FMRFM. This paves the way to imaging of magnetic properties and spin wave dynamics in a variety of contexts for uncovering new physics of nanoscale spin excitations.

## I. Introduction

Localized spin waves are fundamentally important magnetic excitations in ferromagnets (FM) with significant technological implications [1-19]. Ferromagnetic resonance force microscopy (FMRFM) is a powerful spatially-resolved technique for understanding local spin dynamics in buried and exposed magnetic nanostructures with high sensitivity and spectroscopic precision [1-12, 20-23]. FMRFM uses the inhomogeneous magnetic dipolar field of a scanned magnetic probe to create and detect localized spin wave modes [1-3, 6, 8, 9, 12]. This approach offers a unique complement to techniques in which spin waves are localized by the physical boundaries of a patterned structure [4, 5, 10, 18, 19], or through the nonlinear response to a spin polarized current in a nanocontact geometry [13-17]. Utilizing the FMRFM technique to probe the rich spin phenomena in various magnetic materials calls for the ability to control and understand characteristics of the experiment such as localized mode radius and the impact of varying the applied field orientation on spin wave modes. Furthermore, the greater sensitivity of a localized mode to the orientation of local field and magnetization relative to conventional FMR has been rarely discussed. The complexity of the experimental conditions is such that the measurement results cannot be interpreted without the help of micromagnetic modeling. Our modeling results show excellent agreement with the data providing insight into the multiple factors responsible for mode localization, and allowing their response to changing experimental conditions to be tracked. This ability is central to understanding the spin wave physics of the localized modes in various geometries. This method is quite general so its application to a broad range of probes could lead to optimal detection sensitivity and imaging resolution for studying nanoscale magnetic systems.

Currently, the highest-resolution FMRFM localized mode imaging is typically achieved by means of a probe whose magnetic moment is oriented antiparallel to the applied uniform field, creating a region of reduced magnetic field—a well—that confines the spin wave modes directly beneath the probe [1, 6, 8]. This geometry demands a magnetic tip with high coercivity that is time-consuming to fabricate and challenging to create with sub-micron dimensions [1, 6, 8]. An alternative approach in which the spin wave modes are localized by a probe whose moment is parallel to the external field [2-5, 10, 11], and takes advantage of the region of reversed field off the axis of the probe, would eliminate the need for a high coercivity probe and greatly broaden the application of FMRFM by enabling the use of more easily obtainable magnetic probes.

This article reports a systematic study of spin wave modes localized in a $Y_3Fe_5O_{12}$ (YIG) thin film using both parallel and antiparallel geometries that can be quantitatively understood by micromagnetic modelling. Our results demonstrate in-situ tunability of the degree of localization over a broad range by varying the sample-probe separation and the applied field orientation without the need to fabricate patterned structures; this avoids sample imperfections due to edge effect arising from patterning [4, 5, 19]. We predict high spatial resolution in the parallel geometry comparable to the conventional antiparallel geometry. This provides a convenient and versatile method for generating tightly confined localized modes and the potential for high-resolution FMRFM imaging using a wide range of magnetic force microscopy probes, potentially including commercially available cantilevers.

## II. Spin wave mode localization and dynamics in FMRFM

### A. Participating magnetic fields in FMRFM measurements

In our FMRFM experiment a scanned probe with magnetic moment $\boldsymbol{m_p}$ is placed in close proximity to the sample surface. In the general case, an external magnetic field $\boldsymbol{H_0}$ is applied at a small angle $\theta_H$ from the film normal $\hat{n}$ as shown in the insets to Fig. 1. The orientation of $\boldsymbol{m_p}$ is perpendicular to the film plane, either (approximately) be oriented along or opposite to the that of $\boldsymbol{H_0}$, which are referred as "parallel" and "antiparallel" geometries, respectively. The ground state of the position $\boldsymbol{r}$ dependent magnetization $\boldsymbol{M(r)}$ of the YIG film is determined by the total static magnetic field $\boldsymbol{H}_{\text{stat}}(\boldsymbol{r})$ in the film, which is the sum of: (1) the external uniform magnetic field $\boldsymbol{H_0}$, (2) the nonuniform dipolar magnetic field of the probe $\boldsymbol{H}_\text{p}(\boldsymbol{r})$, (3) the nonuniform demagnetizing field of the sample $\boldsymbol{H}_{\text{demag}}(\boldsymbol{r})$, and (4) the effective field describing exchange and anisotropy interactions within the film [24, 25]. In an excited spin wave, the magnetization $\boldsymbol{M(r)}$ undergoes small oscillations about its equilibrium orientation. This can be described as $\boldsymbol{M(r)} = M_s\,\boldsymbol{m(r)}$, where $M_s$ is the saturation magnetization of YIG and $\boldsymbol{m(r)} = \frac{\boldsymbol{M(r)}}{|\boldsymbol{M(r)}|} = \hat{\boldsymbol{x}}\,m_x(\boldsymbol{r}) + \hat{\boldsymbol{y}}\,m_y(\boldsymbol{r}) + \hat{\boldsymbol{z}}\,m_z(\boldsymbol{r})$. We point out that $\hat{\boldsymbol{z}} \parallel \boldsymbol{H}_\text{stat}(\boldsymbol{r})$ and this coordinate frame follows the total static field instead of being fixed to the film geometry. Here $m_x(\boldsymbol{r})$ and $m_y(\boldsymbol{r})$ are the transverse components of magnetization undergoing oscillations about $\boldsymbol{H}_\text{stat}(\boldsymbol{r})$. The various components of $\boldsymbol{H}_\text{stat}(\boldsymbol{r})$, whose inhomogeneity results in spin wave localization, can be broadly tuned by controlling the magnitude and orientation of $\boldsymbol{H_0}$ as well as the strength and spatial profile of $\boldsymbol{H}_\text{p}(\boldsymbol{r})$. This tunability can be achieved by adjusting the probe-sample separation $a$ (see insets to Fig.1) and by selecting "parallel" or "antiparallel" probe configuration at a tilt angle $\theta_H$ of external field, allowing for the study of localized spin wave modes and high-resolution imaging of magnetic dynamics in FMs.

### B. Resonance conditions for localized spin wave modes

It has been demonstrated that the spatial profile of $\boldsymbol{H}_\text{stat}(\boldsymbol{r})$ needed for spin wave localization can occur at sample edges where the inhomogeneity of $\boldsymbol{H}_\text{demag}(\boldsymbol{r})$ dominates [4, 5, 19] or in a region of strongly inhomogeneous probe field $H_p(r)$ [1, 6]. The resonant frequency of the $n^{th}$ localized spin wave mode $\omega_n$ and the spatial profile of the transverse components of magnetization [$m_x(r)$ and $m_y(r)$] are primarily determined by two factors: the spatial profile of $H_{stat}(r)$ and the oscillating magnetic field $h(r) = \hat{x} h_x(r) + \hat{y} h_y(r) + \hat{z} h_z(r)$ created by the precessing magnetization in the localized mode itself. Here $h(r)$ is primarily of magnetic dipolar origin and its spatial profile is determined by the oscillating $m_x(r)$ and $m_y(r)$. The precession frequency $\omega(r)$ of the mode is given by,

$$\left[\frac{\omega(r)}{\gamma}\right]^2 = [H_{stat}(r) - D_{xx}(r)M_s][H_{stat}(r) - D_{yy}(r) M_s], \tag{1}$$

where $\gamma$ is the gyromagnetic ratio, $D_{xx}(r) = \frac{h_x(r)}{m_x(r) M_s}$ and $D_{yy}(r) = \frac{h_y(r)}{m_y(r) M_s}$ are the effective dynamic demagnetizing factors [24] determined by the local $h_x(r)$ and $h_y(r)$ arising from the precessing $m_x(r)$ and $m_y(r)$. Stabilization of the $n^{th}$ localized spin wave mode requires that $\omega(r) = \omega_n$ throughout the region of the localization. However, $H_{stat}(r)$ varies significantly across the mode region mainly due to the contributions from $H_p(r)$ and $H_{demag}(r)$. This requires that the spatial profiles of $D_{xx}(r)$ and $D_{yy}(r)$ adjust accordingly to compensate for the spatial variation of $H_{stat}(r)$ in order to sustain a localized spin wave mode with a constant frequency $\omega(r) = \omega_n$ throughout the mode. For this to happen, the oscillating field $h(r)$ acts effectively as a spatially-varying static magnetic field $H_{dyn}(r)$ that compensates $H_{stat}(r)$. Consequently, Eq. (1) can be written as

$$\left[\frac{\omega(r)}{\gamma}\right]^2 = \left[H_{stat}(r) - H_{dyn}^x(r)\right]\left[H_{stat}(r) - H_{dyn}^y(r)\right], \tag{2}$$

where $H_{dyn}^x(r)$ and $H_{dyn}^y(r)$ are $x$ and $y$ components of $H_{dyn}(r)$. If $D_{xx}(r) = D_{yy}(r) = D(r)$, Eq. (2) can be simplified as [1]

$$H_{\text{eff}} = \frac{\omega_n}{\gamma} = H_{\text{stat}}(r) - H_{\text{dyn}}(r), \tag{3}$$

where $H_{\text{dyn}}(r) = D(r)M_s$ and $H_{\text{eff}}$ is the effective total magnetic field of the mode. Eq. (3) is applicable to any axially symmetric, stable spin wave mode including both uniform and localized modes.

### C. Effects of orientations of applied magnetic field and probe magnetization

We explore the effects of $\boldsymbol{H}_{\text{stat}}(r)$ and $H_{\text{dyn}}(r)$ on spin wave mode localization and frequency, in particular, on how these parameters vary with tunable experimental conditions. Our micromagnetic modelling shows that changes the orientation of the sample magnetization $\boldsymbol{M}(r)$ dramatically affects both $\boldsymbol{H}_{\text{stat}}(r)$ and $H_{\text{dyn}}(r)$. This effect becomes more pronounced when $\boldsymbol{H}_0$ is not orthogonal to the film surface, i.e., $\theta_H > 0°$ (see inset to Fig. 1). In this configuration, the ground state of $\boldsymbol{M}(r)$ is not aligned with $\boldsymbol{H}_0$ due to the strong $\boldsymbol{H}_{\text{demag}}(r)$ in the film. The orientation of $\boldsymbol{M}(r)$ forms an angle $\theta_M > \theta_H$ relative to outward normal to the sample surface $\hat{\boldsymbol{n}}$. Thus, the total static magnetic field in the film can be approximated by [1],

$$\boldsymbol{H}_{\text{stat}}(r) = \boldsymbol{H}_0 + \boldsymbol{H}_{\text{p}}(r) - 4\pi M_s \cos(\theta_M)\hat{\boldsymbol{n}}, \tag{4}$$

where the last term represents the average demagnetizing field $\boldsymbol{H}_{\text{demag}}$ due to the out-of-plane component of $\boldsymbol{M}$. The contributions from anisotropy and exchange [24, 25] to $\boldsymbol{H}_{\text{stat}}(r)$ are not included for clarity. As $\boldsymbol{H}_0$ changes, so does $\theta_M$, which in turn changes $\boldsymbol{H}_{\text{demag}}$ and $\boldsymbol{H}_{\text{stat}}(r)$. In addition, $\boldsymbol{H}_{\text{stat}}(r)$ and $H_{\text{dyn}}(r)$ depend sensitively on the orientation of the probe moment $\boldsymbol{m}_{\text{p}}$ which can be either the "parallel" or "antiparallel" to $\boldsymbol{H}_0$. As a result, the spatially inhomogeneous probe field $\boldsymbol{H}_{\text{p}}(r)$ can either increase or reduce the magnitude of $\boldsymbol{H}_{\text{stat}}(r)$ thus dramatically modifying the conditions for mode localization.

In FMRFM, the strength and orientation of $H_0$ and $m_p$ provide powerful and versatile control "knobs" for manipulation and understanding of the localized spin wave modes. However, it also significantly increases the complexity of the experimental configuration and makes it challenging to interpret the observed results. This is why analytical calculation of mode dynamics and localization has only been used successfully in high symmetry situations [26-28]. To fully take advantage of the versatility of FMRFM in various configurations, we employ numeric micromagnetic modelling to interpret our results and understand the localized spin wave dynamics.

### III. Sample and probe preparation

We use a 25-nm thick YIG epitaxial thin film grown by off-axis sputtering [29-31] on a (111)-$Gd_3Ga_5O_{12}$ (GGG) substrate for FMRFM experiment. YIG has attracted a great deal of attention in spin wave [32, 33], spin transport, and spin dynamics [6, 34-39] studies due to its exceptionally low damping, small coercivity, moderate saturation magnetization and high efficiency of angular momentum transfer [6, 35-38]. The YIG film with a saturation magnetization of $4\pi M_s = 1592$ Oe is cut into a strip of approximately $5 \times 2$ mm$^2$ and glued on a microwave transmission line. Our FMRFM probe uses a $SmCo_5$ magnetic particle of 1.74 $\mu$m in diameter with a magnetic moment of $3.9 \times 10^{-9}$ emu and coercivity of 10000 Oe measured by cantilever magnetometry [40] is glued at the end of a commercial cantilever [1, 6]. FMRFM signal is obtained by measuring the cantilever amplitude as a function of $H_0$ at a fixed radio-frequency (rf) $f_{rf} = 2.157$ GHz. To improve detection sensitivity, the amplitude of the output microwave power is modulated at the resonance frequency of the cantilever (~18 kHz).

### IV. Micromagnetic Modelling

The micromagnetic modelling employed custom modelling software developed at The Ohio State University using MATLAB®, the high-level language for technical computing. The thin film sample is approximated by a 2D array of thin, uniformly magnetized prisms. The magnetization dynamics in a prism is described by a linearized Landau-Lifshitz-Gilbert equation which includes the interaction of the prism's magnetization with the external magnetic field and the effective field which describes interactions with the other prisms in the array. Such an equation is written for each prism in the array thus resulting in a system of linear equations. The resonant fields and the spatial profiles of the modes are obtained by finding the eigenvalues and the eigenstates of this system of equations using numerical solvers provided by MATLAB®. To reduce calculation time, a variable mesh grid is used such that ~ 900 to 6400 small prisms (lateral dimensions as small as $10\times10$ nm$^2$) are enclosed within the localized mode region under study, while areas outside the mode are approximated by larger prisms. The calculations are repeated for several grid choices to verify that the calculated results do not change with grid size.

## V. Experimental results

Figure 1 shows a series of FMRFM spectra recorded for multiple probe-sample separations $a$ in both "antiparallel" [Fig. 1(a)] and "parallel" [Fig. 1(b)] configurations at $\theta_\mathrm{H} = 0°$ ($\boldsymbol{H}_0 \perp$ film plane). All the spectra show a similar feature at $H_0 = 2357$ Oe independent of $a$, which is attributed to the resonance of the uniform mode [1, 6, 8, 9]; this field at which the uniform mode is in resonance is designated as $H_0^\mathrm{unif}$ for further discussion. The features at $H_0 > H_0^\mathrm{unif}$ are attributed to the spin wave modes localized by the probe field $\boldsymbol{H}_\mathrm{p}(\boldsymbol{r})$. The mode at the highest field is the 1$^\mathrm{st}$ localized mode [1, 6] and the corresponding field is designated as $H_0^\mathrm{loc}$. The spectra in Figs. 1(a) and 1(b), however, demonstrate a striking difference in localized mode formation between the "antiparallel" and "parallel" configurations. In the "antiparallel" case, localized modes clearly appear at $a$ = 4850 nm and the field shift between the 1$^{\text{st}}$ localized and the uniform mode $H_0^{\text{loc}} - H_0^{\text{unif}}$ increases rapidly to 213 Oe at $a$ = 2250 nm. In contrast, for the "parallel" configuration the localized mode does not appear until the probe is brought within 1000 nm of the film surface and the mode shift $H_0^{\text{loc}} - H_0^{\text{unif}}$ is much smaller, e.g., 30 Oe at $a$ = 190 nm. This arises from the significant difference in the profiles of $\boldsymbol{H}_{\text{stat}}(\boldsymbol{r})$ and $H_{\text{dyn}}(\boldsymbol{r})$ in the two configurations.

To probe the influence of $\boldsymbol{H}_0$ orientation on the resonance condition of the localized modes, we tilt $\boldsymbol{H}_0$ away from film normal in both "antiparallel" and "parallel" configurations, as shown in Fig. 2a for $\theta_{\text{H}}$ = 0°, 4°, and 6°. The spectra are plotted vs $H_0^{\text{loc}} - H_0^{\text{unif}}$ for clarity due to the changes of $H_0^{\text{loc}}$ [inset to Fig. 2(a)] and $H_0^{\text{unif}}$ with $\theta_{\text{H}}$. This tilting of field direction results in variation of the shift $H_0^{\text{loc}} - H_0^{\text{unif}}$ and particularly affects the localized modes in the "parallel" configuration more profoundly. In the "parallel" configuration, $H_0^{\text{loc}} - H_0^{\text{unif}}$ increases by 151 Oe as $\theta_{\text{H}}$ increases from 0° to 6° while $H_0^{\text{loc}} - H_0^{\text{unif}}$ increases by only 47 Oe in the "antiparallel" case. Figure 2(b) summarizes the dependence of $H_0^{\text{loc}} - H_0^{\text{unif}}$ on probe-sample separation $a$ and tilt angle $\theta_{\text{H}}$ for both configurations, where the symbols are the experimental data points and the solid curves are the results of micromagnetic modelling as discussed below. This figure reveals that while $H_0^{\text{loc}} - H_0^{\text{unif}}$ in the "antiparallel" case is more sensitive to the probe-sample separation, the "parallel" configuration exhibits a much stronger dependence on the tilt angle $\theta_{\text{H}}$. For example, in the parallel case, as $\boldsymbol{H}_0$ tilts from $\theta_{\text{H}}$ = 0° to 6°, $H_0^{\text{loc}} - H_0^{\text{unif}}$ increases dramatically from 15 Oe to 199 Oe at $a$ = 660 nm, which is close to the shift for "antiparallel" configuration at separation of 2250 nm and $\theta_{\text{H}}$ = 0°. This high sensitivity of localized spin wave modes to a small tilt angle of $\boldsymbol{H}_0$ implies broad tunability in controlling nanoscale spin dynamics using the less frequently used "parallel" geometry. The excellent agreement between the experimental data and the micromagnetic modelling allows us to extract essential parameters of the localized modes. As an example, we show in Fig. 2(c) the characteristic dimensions of the 1$^{st}$ localized mode extracted from micromagnetic modeling for the experimental data presented in Fig. 2b. Figure 2(d) and (e) show the 3D dependence of characteristic mode size as a function of probe-sample separation and angle in the "parallel" and "antiparallel" geometries respectively. The size of the localized mode decreases as the probe is brought closer to the film surface. We note that the "parallel" configuration exhibits more significant reduction in mode radius with decreasing $a$ and higher sensitivity to $\theta_H$, suggesting that the "parallel" orientation can be used for sensitive control of mode localization and to achieve imaging resolution comparable to the "antiparallel" case.

## VI. Micromagnetic modelling of spin wave mode localization in various geometries

Micromagnetic modelling enables detailed analysis of the various parameters describing magnetization dynamics, allowing these parameters to be tracked as the magnitude and direction of $\boldsymbol{H}_0$ are varied. As we have demonstrated earlier [1], the localized modes are confined in the region where the total static field $\boldsymbol{H}_{\text{stat}}(\boldsymbol{r})$ forms a field "well" relative to the rest of the sample produced by the probe field $\boldsymbol{H}_{\text{p}}(\boldsymbol{r})$. Our micromagnetic simulations indicate that the "well" occurs directly below the probe in the "antiparallel" configuration [inset to Fig. 1(a) and Fig. 3(a)] and to the sides of the probe in the "parallel" configuration [inset to Fig. 1(b) and Fig.3(d)]. Meanwhile, the uniform mode forms in the regions far from the probe, where $H_{\text{p}}(\boldsymbol{r}) \approx 0$; thus, its resonant field is independent of the probe-sample separation as shown in Fig. 1. The spatial dependence of $\boldsymbol{H}_{\text{stat}}(\boldsymbol{r})$ can be divided into three distinct regions where different approximations apply,

$$H_{\text{stat}}(r) = \begin{cases} H_{\text{stat}}^{\text{loc}}(r), & \text{region where the localized mode is stable} \\ H_{\text{stat}}^{\text{unif}}, & \text{region where the uniform mode is stable} \\ H_{\text{stat}}^{\text{none}}(r), & \text{region where neither mode is stable} \end{cases} \quad (5)$$

The static field in the region of the uniform spin wave mode (away from the FMRFM probe) is essentially constant and Eq. (4) can be approximated by

$$H_{\text{stat}}^{\text{unif}} = H_0 - 4\pi M_s \cos(\theta_M)\hat{n} = H_0 + H_{\text{demag}}^{\text{unif}}, \quad (6)$$

where $H_{\text{demag}}^{\text{unif}}$ is the static demagnetizing field in the region of the sample where the uniform mode is stable and $\theta_M$ can be determined by [24, 25]:

$$\tan(\theta_M) = \frac{H_0 \sin(\theta_H)}{H_0 \cos(\theta_H) - 4\pi M_s \cos(\theta_M)}, \quad (7)$$

which implies that $\theta_M > \theta_H$ if $\theta_H > 0°$. As $H_0$ increases, $\theta_M$ becomes smaller and approaches $\theta_H$, thus making $H_{\text{demag}}^{\text{unif}}$ more negative and reducing $H_{\text{stat}}^{\text{unif}}$.

$H_{\text{stat}}(r)$ in the region of the localized mode is significantly more complicated due to the presence of the strongly inhomogeneous probe field $H_p(r)$ and Eq. (4) can be rewritten as

$$H_{\text{stat}}^{\text{loc}}(r) = H_0 + H_{\text{demag}}(r) + H_p(r). \quad (8)$$

There is no analytical approximation describing $H_{\text{demag}}(r)$ in this case, which makes micromagnetic modeling an indispensable tool for analyzing the problem. Figure 3 shows our calculated spatial profiles of the out-of-plane components of $H_{\text{demag}}(r)$ and $H_p(r)$ as well as the total static field $H_{\text{stat}}(r)$ across the region under the probe for the "antiparallel" and "parallel" configurations at $\theta_H = 0°$ and $6°$. The ability to visualize the spatial profiles of individual contributions to $H_{\text{stat}}(r)$ offers insight into the evolution of critical parameters that determine the localized spin wave dynamics created and probed by FMRFM.

We first discuss the impact of field tilting on $H_{\text{demag}}(r)$. As can be seen in Figs. 3(a) and 3(d) for $\theta_H = 0°$, $H_{\text{demag}}(r)$ is symmetric with a small magnitude of variations of ~10 Oe at $a = 2500$ nm for the "antiparallel" case and ~100 Oe at $a = 1000$ nm for the "parallel" case. The two peaks directly beneath the probe stem from a moderate tilt of $M(r)$ relative to $\hat{n}$ as schematically indicated in the insets to Figs. 3(a) and 3(d) which is caused by the presence of the probe field $H_p(r)$. At $\theta_H = 6°$, $H_{\text{demag}}(r)$ becomes more asymmetric and the spatial variation is significantly larger, as shown in Figs. 3(b) and 3(e).

Meanwhile, the probe field $H_p(r)$ is independent of $\theta_H$, but its profile is dramatically different between the "antiparallel" and "parallel" configurations. For the "antiparallel" case [Figs. 3(a) and 3(b)], the strong negative $H_p(r)$ creates a deep field "well" directly beneath the probe. Since $H_p(r)$ is significantly stronger than the variation of $H_{\text{demag}}(r)$, the field "well" of $H_{\text{stat}}^{\text{loc}}(r)$ [Fig. 3(c)] only changes slightly with $\theta_H$. As a result, the localized spin wave modes in the "antiparallel" configuration show a weak dependence on $\theta_H$. In contrast, $H_p(r)$ exhibits a dominant peak (~1000 Oe) in the "parallel" configuration with a shallow field "well" (~20 Oe) 1 μm away from the probe location due to the dipolar nature of $H_p(r)$. The weak field "well" in the "parallel" case explains why a noticeable shift of the localized modes requires much closer probe-sample separation compared to the "antiparallel" configuration (Fig. 1). Since the depth of the side "well" of $H_p(r)$ is comparable to the variations of $H_{\text{demag}}(r)$ in the "parallel" configuration, tilting of $H_0$ can significantly modify the overall "well" of $H_{\text{stat}}^{\text{loc}}(r)$. This has a profound effect on the formation of localized spin wave modes as shown in Fig. 3(f) in which the field "well" to the left of the peak becomes much deeper at $\theta_H = 6°$ as compared to that at $\theta_H =$ 0°, demonstrating broad-range tunability of localized modes by controlling the orientation of $\boldsymbol{H}_0$ using the "parallel" geometry.

In order to further understand the $\theta_H$ dependence of $H_0^{loc} - H_0^{unif}$ as shown in Fig. 2(a), we discuss the influence of $H_{dyn}(\boldsymbol{r})$ introduced in Eq. (2) on the resonant properties of a spin wave mode. As shown by Eq. (3), the effective field $H_{eff}$ of a mode has both static and dynamic field contributions. In general, the dependence of $H_{eff}$ on $H_{dyn}(\boldsymbol{r})$ is complicated given that $H_{dyn}(\boldsymbol{r})$ has two orthogonal components [Eq. (2)] determined by the geometry. Because of $H_{dyn}$, $H_{eff}$ for a stable spin wave mode is usually greater or equal to the minimum value of $H_{stat}(\boldsymbol{r})$ in the region of the mode [see Eq. (3)]. We define the peak effective dynamic field of the localized and uniform modes as $H_{dyn}^{loc} = H_{eff}^{loc} - \min[H_{stat}^{loc}(\boldsymbol{r})]$ and $H_{dyn}^{unif} = H_{eff}^{unif} - \min[H_{stat}^{unif}(\boldsymbol{r})]$, respectively.

Figure 4 shows a numerical calculation of the contributions of $H_{dyn}^{loc}$ and $H_{dyn}^{unif}$ to the effective resonant field of the localized and uniform modes for $\theta_H = 0°$ and $6°$ at $a = 1000$ nm in the "parallel" configuration, which compares the values of $H_{eff}^{unif}$ and $H_{eff}^{loc}$ with the corresponding $H_{stat}(\boldsymbol{r})$ profiles. The external field $H_0$ is set to $H_0^{unif}$ at which the uniform mode for a given $\theta_H$ is resonant with the effective rf field $\omega_{rf}/\gamma$. Both the static field profile $\boldsymbol{H}_{stat}(\boldsymbol{r})$ and the peak effective dynamic fields $H_{dyn}^{loc}$ and $H_{dyn}^{unif}$ change significantly with $\theta_H$. The profile of $H_{stat}(\boldsymbol{r})$ becomes asymmetric with a deeper and narrower field "well" at $\theta_H = 6°$, which localizes modes.

The significant increase of $H_{dyn}^{loc}$ and $H_{dyn}^{unif}$ with increasing $\theta_H$ originates in part from the change of orientation of the sample magnetization $\boldsymbol{M}(\boldsymbol{r})$ within the extent of the mode. The change in $\theta_M$ also alters the orientation of the oscillation plane of the transverse magnetization $m_x(\mathbf{r})$ and $m_y(\mathbf{r})$ relative to the film surface. At $\theta_H = 0°$, this oscillation plane is parallel to the film surface; thus, for the uniform mode, the oscillating transverse magnetization does not have components normal to the film surface, resulting in zero effective magnetic charge density on the film surface [41] and $H_{\text{dyn}}^{\text{unif}} = 0$. With increasing $\theta_H$, this charge density starts to grow since the component of oscillating magnetization normal to sample surface becomes nonzero. This larger effective surface charge density increases the strength of the oscillating magnetic field $\mathbf{h}(\mathbf{r})$, thus increasing the strength of $H_{\text{dyn}}(\mathbf{r})$. This effect is particularly pronounced for the uniform mode as shown in Fig. 4, where the peak effective dynamic field $H_{\text{dyn}}^{\text{unif}} \approx 0$ Oe at $\theta_H = 0°$ and increases to $H_{\text{dyn}}^{\text{unif}} \approx 83$ Oe at $\theta_H = 6°$. The increase of the peak dynamic field of the localized mode $H_{\text{dyn}}^{\text{loc}}$ at $\theta_H = 6°$ stems from a narrower field "well". As discussed previously [1], the narrower confinement of a mode results in a stronger effective dynamic field $H_{\text{dyn}}(\mathbf{r})$ due to the closer effective magnetic charges formed at the edges of the mode.

The combined changes in static and dynamic fields with increasing $\theta_H$ result in significant increase in the shift $H_{\text{eff}}^{\text{unif}} - H_{\text{eff}}^{\text{loc}}$ from 9 Oe at $\theta_H = 0°$ to 153 Oe at $\theta_H = 6°$ (Fig. 4) in the "parallel" configuration calculated from micromagnetic simulation. The shift in $H_{\text{eff}}$ manifests itself as $H_0^{\text{loc}} - H_0^{\text{unif}} = 159$ Oe in the experiment at $\theta_H = 6°$ [Fig. 2(b)], which is very close to our calculated value of $H_{\text{eff}}^{\text{unif}} - H_{\text{eff}}^{\text{loc}}$ at the same $\theta_H$. Our simulations reveal why these two values are nearly equal: the transverse field arising from the probe partially compensates the in-plane component of the tilted applied field. Thus the average sample magnetization is coincidentally very nearly aligned with the applied field within the localization region. This "self-correcting" feature is attractive for experiments as it reduces artificial enhancement to the linewidth that can be experienced with off-axis magnetic fields, indicating the possibility of using the localized mode generated in the "parallel" configuration for linewidth analysis to reveal new physics [6, 19]. The excellent agreement between the experiment and the numerical results as shown in Fig. 2(b) lends credibility to the numeric model that we use and enables understanding of the mode dependence on the parameters of the experiment.

## VII. Control of spin wave mode localization in "parallel" probe configuration

Micromagnetic modelling allows us to visualize the spatial profiles of the localized modes encountered in the experiment in both "parallel" and "antiparallel" geometries. This leads to a central finding of this article that is shown in Fig. 5 which highlights the spatial profile, modulus of the transverse component of the dynamic magnetization, of the 1$^{st}$ localized mode calculated for both "antiparallel" and "parallel" configurations and its $\theta_H$ dependence. The lateral size of the mode in the "antiparallel" configuration is nearly insensitive to $\theta_H$ and its location only shifts slightly relative to the probe [Fig. 5(a)] because the field "well" localizing the mode does not change significantly with $\theta_H$ as shown in Fig. 3(c). In contrast, the localized mode in the "parallel" configuration reduces its size dramatically, from 15 μm × 15 μm down to 1 μm × 2 μm as $\theta_H$ increases from 0° to 6° [Fig. 5(b)]. The shape of the mode changes from a "doughnut" shape at 0° to a much smaller dot-like shape at 6° while the mode location also shifts to the side of the probe. This change stems from the deep asymmetry of the localizing field "well" in the "parallel" configuration induced by the tilting of $\boldsymbol{H}_0$ as shown in Fig. 3(f) and Fig. 4. This powerful and convenient method for controlling the mode confinement in the "parallel" configuration signifies a major advance for the FMRFM technique.

Previously, typical localized mode imaging FMRFM experiments were conducted using micromagnetic probe in the "antiparallel" geometry which resulted in highly confined localized modes with characteristic radii as low as 200 nm [1]. This configuration requires fabrication of a custom high coercivity magnetic probe with a magnetic moment $m_p$ that does not get reversed by the opposing external magnetic field $H_0$. This was typically achieved by gluing a high coercivity (10000 – 15000 Oe) magnetic particle to the end of a commercial Atomic Force Microscopy (AFM) cantilever followed by focused-ion-beam milling of the particle to a characteristic dimension of ~1 μm, a complicated and labor intensive process [1, 6, 8]. Commercial MFM cantilevers cannot be used in the "antiparallel" configuration due to their relatively low coercivity (~1000 Oe). However, the "parallel" configuration does not require a high coercivity probe since $m_p \parallel H_0$, offering a new path for using commercial MFM cantilevers in FMRFM experiments.

To evaluate this new approach, we perform micromagnetic calculations of mode localization in the "parallel" configuration at an rf frequency of 2.157 GHz for a typical MFM probe approximated by a magnetic sphere of 50-nm radius [42, 43] and $4\pi M_s$ = 15000 Oe. Figure 6 shows the probe-sample separation dependence of the radius of the 1$^{st}$ spin wave mode in a 25 nm YIG thin film generated by a commercial MFM cantilever in the "parallel" configuration at $\theta_H = 9°$ ($m_p \parallel H_0$). The resulting elliptical mode shape is characterized by $R_{long}$ and $R_{short}$ radii as indicated in the inset showing the localized mode shape calculated for $a$ = 10 nm. The radius of the mode was estimated using the 10% of the peak mode amplitude rule [1, 2]. The localized mode size decreases with reducing probe-sample separation, thus increasing achievable spatial resolution of FMRFM imaging. The results show that a spatial resolution of $R_{short}$ = 123 nm and $R_{long}$ = 211 nm in YIG is achievable at 10 nm probe-sample separation. The expected imaging resolution is similar to that achieved with a custom probe [1, 2], suggesting that the use of MFM cantilevers with soft magnetic coating [5, 44] for high resolution FMRFM imaging is plausible and promising.

## VIII. Conclusion

In conclusion, we demonstrate a broadly tunable approach to generating localized spin wave modes in magnetic materials using FMRFM. The resonance field, spatial profile, position of the localization, and mode size can be systematically tuned by controlling the orientation of applied uniform field relative to the sample plane and probe moment. Our micromagnetic modeling accurately reproduces the observed experimental results and enables understanding of the localized spin wave dynamics in a wide range of configurations. This provides a universal method to understand and control the characteristics of localized spin wave modes, which is fundamentally important for the study of static and dynamic spin properties in a variety of nanoscale systems. The ability to use a wider variety of micromagnetic probes to create tightly confined spin wave modes for high-resolution FMRFM imaging will improve accessibility and ease application of FMRFM.

**Acknowledgments**

This work was primarily supported by the U.S. Department of Energy (DOE), Office of Science, Basic Energy Sciences, under Award No. DE-FG02-03ER46054 (FMRFM characterization and modelling) and No. DE-SC0001304 (sample synthesis). This work was supported in part by the Center for Emergent Materials, an NSF-funded MRSEC under award No. DMR-1420451 (structural and magnetic characterization). This work was supported in part by an allocation of computing time from the Ohio Supercomputer Center. We also acknowledge technical support and assistance provided by the NanoSystems Laboratory at the Ohio State University. Partial support was provided by Lake Shore Cryogenics, Inc.

21. I. Lee, Y. Obukhov, J. Kim, X. Li, N. Samarth, D. V. Pelekhov, and P. C. Hammel, *Local magnetic characterization of (Ga,Mn)As continuous thin film using scanning probe force microscopy*, Phys. Rev. B **85**, 184402 (2012).

22. K. Wago, D. Botkin, C. S. Yannoni, and D. Rugar, *Paramagnetic and ferromagnetic resonance imaging with a tip-on-cantilever magnetic resonance force microscope*, Appl. Phys. Lett. **72**, 2757 (1998).

23. E. Nazaretski, I. Martin, R. Movshovich, D. V. Pelekhov, P. C. Hammel, M. Zalalutdinov, J. W. Baldwin, B. Houston, and T. Mewes, *Ferromagnetic resonance force microscopy on a thin permalloy film*, Appl. Phys. Lett. **90**, 234105 (2007).

24. S. V. Vonsovskii, *Ferromagnetic Resonance* (Pergamon, Oxford, 1966).

25. X. Liu, W. L. Lim, L. V. Titova, M. Dobrowolska, J. K. Furdyna, M. Kutrowski, and T. Wojtowicz, *Perpendicular magnetization reversal, magnetic anisotropy, multistep spin switching, and domain nucleation and expansion in $Ga_{1-x}Mn_xAs$ films*, J. Appl. Phys. **98**, 063904 (2005).

26. G. N. Kakazei, P. E. Wigen, K. Yu. Guslienko, V. Novosad, A. N. Slavin, V. O. Golub, N. A. Lesnik, and Y. Otani, *Spin-wave spectra of perpendicularly magnetized circular submicron dot arrays*, Appl. Phys. Lett. **85**, 443 (2004).

27. V. Castel, J. Ben Youssef, F. Boust, R. Weil, B. Pigeau, G. de Loubens, V. V. Naletov, O. Klein, and N. Vukadinovic, *Perpendicular ferromagnetic resonance in soft cylindrical elements: Vortex and saturated states*, Phys. Rev. B **85**, 184419 (2012).
21

**Table I**. Definition of variables used in our calculations and discussion,

| | |
|---|---|
| $H_0$ | External applied field |
| $a$ | Probe-sample separation |
| $\omega$ | Precession frequency of mode |
| $\hat{n}$ | Normal of the film plane |
| $\theta_H$ | Tilted angle of external field relative to the normal of the film plane |
| $M$ | Magnetization |
| $\hat{m}$ | Unit vector describing orientation of magnetization as described in the text |
| $4\pi M_s$ | Saturation magnetization |
| $h$ | Oscillating magnetic field |
| $\theta_M$ | Tilted angle of magnetization to the normal of the film plane |
| $H_0^{unif}$ | External field at which uniform mode is in resonance |
| $H_0^{loc}$ | External field at which 1$^{st}$ localized mode is in resonance |
| $H_p$ | Probe field, spatially varying dipole field from the magnetic particle on cantilever |
| $m_p$ | Magnetic moment of probe |
| $H_{dyn}$ | Effective static magnetic field describing the effects of dynamic magnetic fields in the system ( as describe in text) |
| $H_{demag}$ | Demagnetizing magnetic field. In thin film samples $H_{demag}$ is primarily due to the out of plane component of static magnetization |
| $D$ | Effective dynamic demagnetizing factor as described in the text |
| $H_{dyn}^{unif}$ | Peak effective dynamic field of uniform mode as described in the text: $H_{dyn}^{unif} = H_{eff}^{unif} - \min(H_{stat}^{unif}(\boldsymbol{r}))$ |
| $H_{dyn}^{loc}$ | Peak effective dynamic field of uniform mode as described in the text: $H_{dyn}^{loc} = H_{eff}^{loc} - \min(H_{stat}^{loc}(\boldsymbol{r}))$ |
| $H_{\text{dyn}}^{\text{unif}}(0°)$ | Peak effective dynamic field of uniform mode evaluated at $\theta_H = 0°$ |
| $H_{\text{dyn}}^{\text{unif}}(6°)$ | Peak effective dynamic field of uniform mode evaluated at $\theta_H = 6°$ |
| $H_{\text{dyn}}^{\text{loc}}(0°)$ | Peak effective dynamic field of 1$^{\text{st}}$ localized mode evaluated at $\theta_H = 0°$ |
| $H_{\text{dyn}}^{\text{loc}}(6°)$ | Peak effective dynamic field of 1$^{\text{st}}$ localized mode evaluated at $\theta_H = 6°$ |
| $H_{\text{stat}}$ | Magnitude of the total static field including contributions of $H_0$, $H_{\text{demag}}$ and $H_p$ |
| $H_{\text{stat}}^{\text{unif}}(0°)$ | $H_{\text{stat}}$ of uniform mode evaluated at $\theta_H = 0°$ |
| $H_{\text{stat}}^{\text{unif}}(6°)$ | $H_{\text{stat}}$ of uniform mode evaluated at $\theta_H = 6°$ |
| $\omega_{\text{rf}}/\gamma$ | Effective field of microwave frequency |
| $H_{\text{eff}}^{\text{unif}}$ | Effective field $\omega/\gamma$ of uniform mode as described in text |
| $H_{\text{eff}}^{\text{loc}}$ | Effective field $\omega/\gamma$ of 1$^{\text{st}}$ localized mode as described in text |
| $H_{\text{eff}}^{\text{unif}}(0°)$ | Effective field $\omega/\gamma$ of uniform mode evaluated at $\theta_H = 0°$ |
| $H_{\text{eff}}^{\text{unif}}(6°)$ | Effective field $\omega/\gamma$ of uniform mode evaluated at $\theta_H = 6°$ |
| $H_{\text{eff}}^{\text{loc}}(0°)$ | Effective field $\omega/\gamma$ of 1$^{\text{st}}$ uniform localized mode evaluated at $\theta_H = 0°$ |
| $H_{\text{eff}}^{\text{loc}}(6°)$ | Effective field $\omega/\gamma$ of 1$^{\text{st}}$ uniform localized mode evaluated at $\theta_H = 6°$ |

**Figure Captions:**

**Figure 1.** FMRFM spectra taken at $f_{rf}$ = 2.157 GHz and at various probe-sample separations $a$ when the probe magnetic moment $m_p$ is (a) "antiparallel" and (b) "parallel" to the applied uniform field $H_0$ at $\theta_H = 0°$ ($H_0$ normal to the film plane). Insets: schematics of experimental configurations of FMRFM measurements, where the black and green curves represent the spatial profiles of the magnitude of probe field $H_p$ and the amplitude of the 1st localized mode. Spectra are offset for clarity.

**Figure 2.** (a) Selected FMRFM spectra as a function of $H_0 - H_0^{unif}$ at $\theta_H = 0°$, 4°, and 6° for both "parallel" ($a$ = 1000 nm) and "antiparallel" ($a$ = 2500 nm) configurations. Inset: resonance field $H_0^{loc}$ of the 1st localized mode as a function of $a$ at $\theta_H = 0°$ and 4° in the "antiparallel" geometry. (b) Shift in resonance field between the 1st localized mode and the uniform mode $H_0^{loc} - H_0^{unif}$ as a function of $a$ at various $\theta_H$ in the "antiparallel" (solid squares) and "parallel" (solid circles) configurations. The solid curves are micromagnetic modelling results which agree well with the experimental data. (c) Characteristic size of the 1st localized mode at various $\theta_H$ in the "antiparallel" and "parallel" configurations obtained by micromagnetic modeling, which represents the radius of the short axis of the mode in different measurement geometries (for $\theta_H$ = 0° in the "parallel" configuration with a "doughnut" shape mode, we use the difference between the outer and inner radius, see Fig. 5). 3D dependences of characteristic size of the 1st localized mode as a function of probe-sample separation and angle for (d) the "parallel" and (e) "antiparallel" geometries, respectively, emphasizing the strong contrast of the angular dependencies of characteristic mode sizes.

**Figure 3.** Spatial variation of the out-of-plane components of $H_{demag}(r)$ and $H_p(r)$ at (a) $\theta_H$ = 0° and (b) $\theta_H$ = 6° as well as (c) $H_{stat}(r)$ for both angles in the "antiparallel" configuration.

Corresponding plots for the "parallel" configuration are shown in (d), (e), and (f). Insets: schematics of the spatial profiles of the equilibrium orientation of magnetization $M$ in that particular configuration. Note the two y-axes in each plot are offset relative to each other for comparison. It demonstrates that the spatial profile of $H_{\text{stat}}(r)$ in the "antiparallel" configuration in (c) only changes slightly with $\theta_H$ while the depth of the field "well" in $H_{\text{stat}}(r)$ in the "parallel" case in (f) changes significantly.

**Figure 4.** Comparison of $H_{\text{eff}}^{\text{unif}}$ and $H_{\text{eff}}^{\text{loc}}$ with the corresponding $H_{\text{stat}}(r)$ numerically calculated for $\theta_H = 0°$ and $6°$ at probe-sample separation $a = 1000$ nm in the "parallel" configuration. The external field is set to $H_0^{\text{unif}}$, the field at which the uniform mode is resonant with the effective rf field $\omega_{\text{rf}}/\gamma$. The static field profile $H_{\text{stat}}(r)$, peak effective dynamic fields $H_{\text{dyn}}^{\text{loc}}$, and $H_{\text{dyn}}^{\text{unif}}$ change significantly with $\theta_H$. For example, $H_{\text{dyn}}^{\text{unif}}(0°) \approx 0$ at $\theta_H = 0°$ changes to $H_{\text{dyn}}^{\text{unif}}(6°) \approx 83$ Oe at $\theta_H = 6°$. These changes are the origin of the strong angular dependence of the localized mode resonance in the "parallel" configuration observed in the experiment.

**Figure 5.** Spatial profile, modulus of the transverse component of the dynamic magnetization, of the 1st localized mode calculated for (a) the "antiparallel" ($a = 2500$ nm) and (b) "parallel" ($a = 1000$ nm) configurations for $\theta_H = 0°$ and $6°$. The lateral size of the mode in the "antiparallel" case essentially remains unchanged with $\theta_H$, while mode size in the "parallel" configuration is greatly reduced as $\theta_H$ increases from $0°$ to $6°$. For all panels the probe is located at (0, 0).

**Figure 6.** Micromagnetic modelling of the characteristic dimensions of the 1st localized spin wave mode as a function of the probe-sample separation in a 25-nm YIG thin film at $\theta_H = 9°$ generated by a commercial MFM cantilever in the "parallel" configuration. Variable density mesh approximation with the cell size as small as 10×10 nm was used for simulations. The rf frequency is 2.157 GHz and the MFM probe is assumed to be a magnetic sphere with a radius of 50 nm and saturation magnetization $4\pi M_s$ = 15000 Oe. The elliptical mode shape is characterized by the long ($R_{long}$) and short ($R_{short}$) radii as indicated in the inset calculated for 10 nm probe-sample separation.

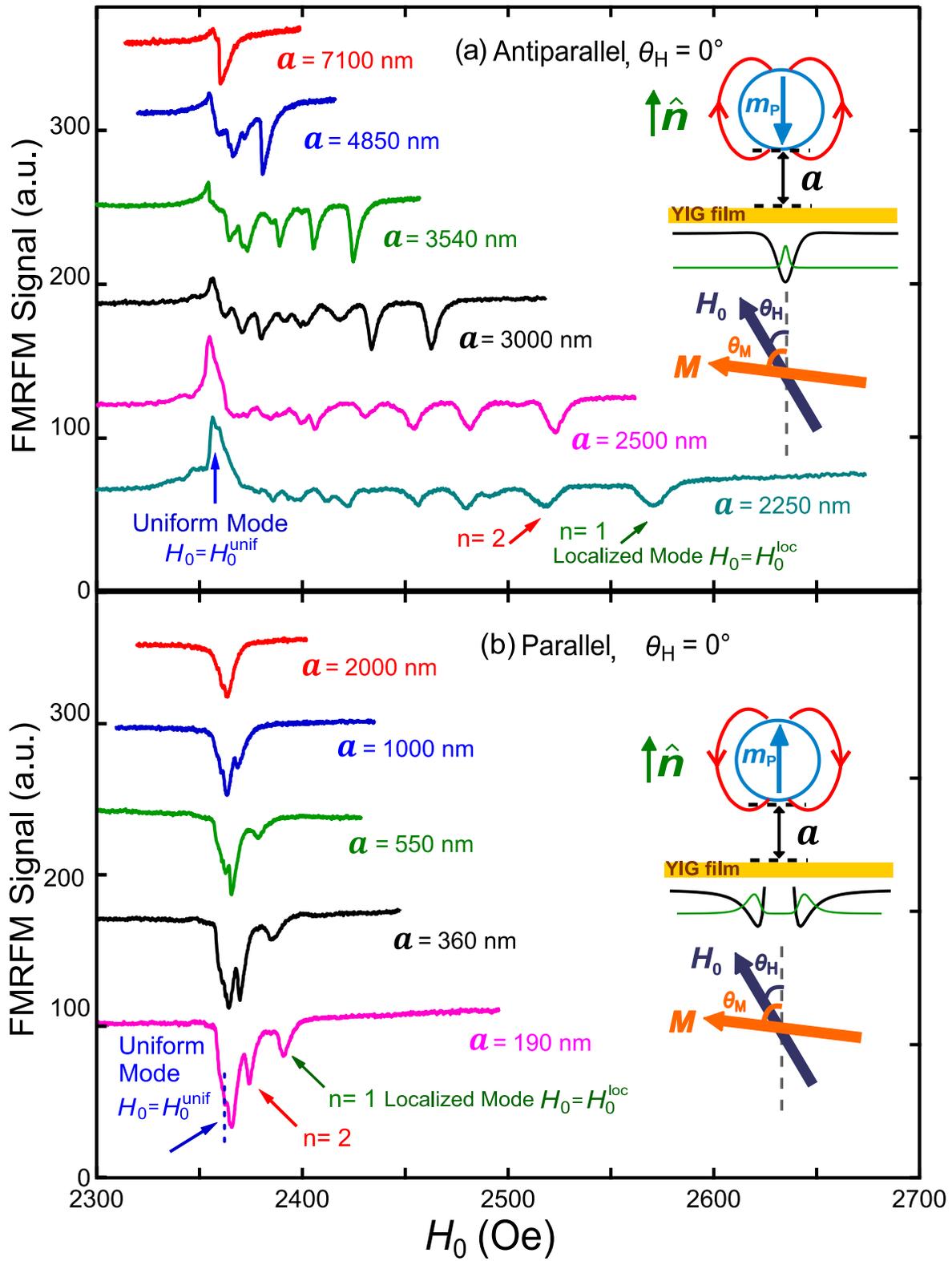

**Figure 1.**

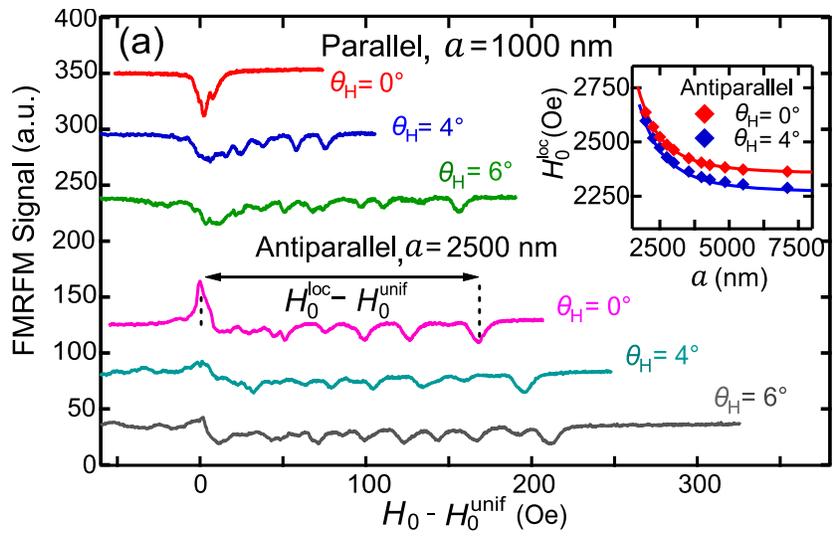
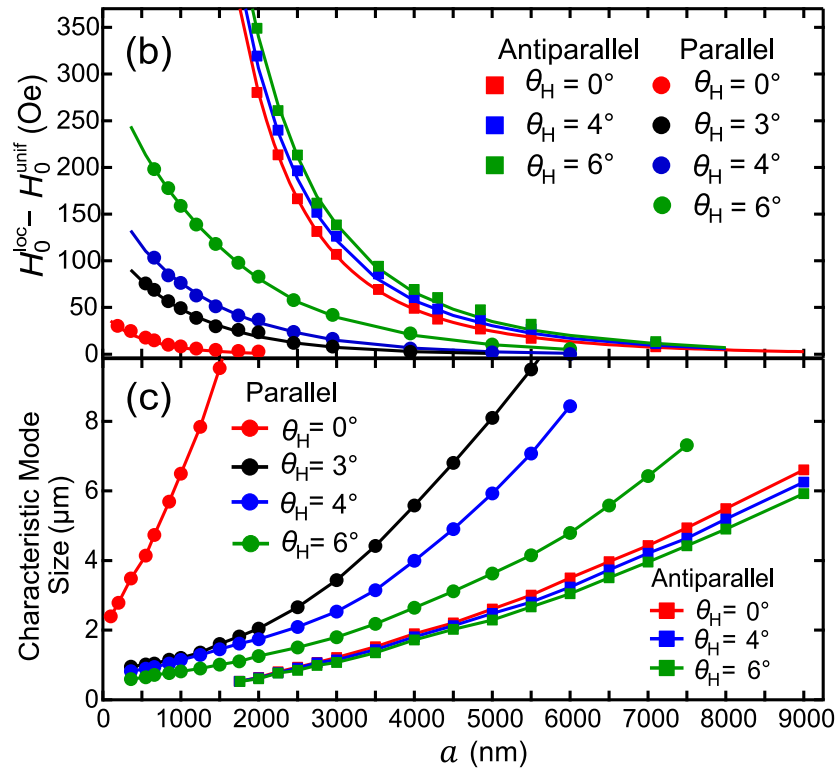
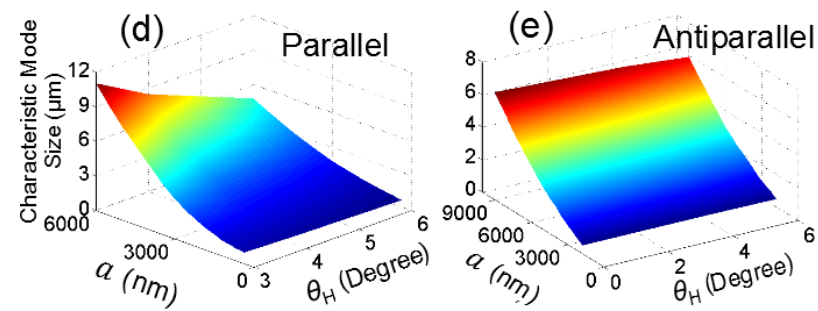

**Figure 2**

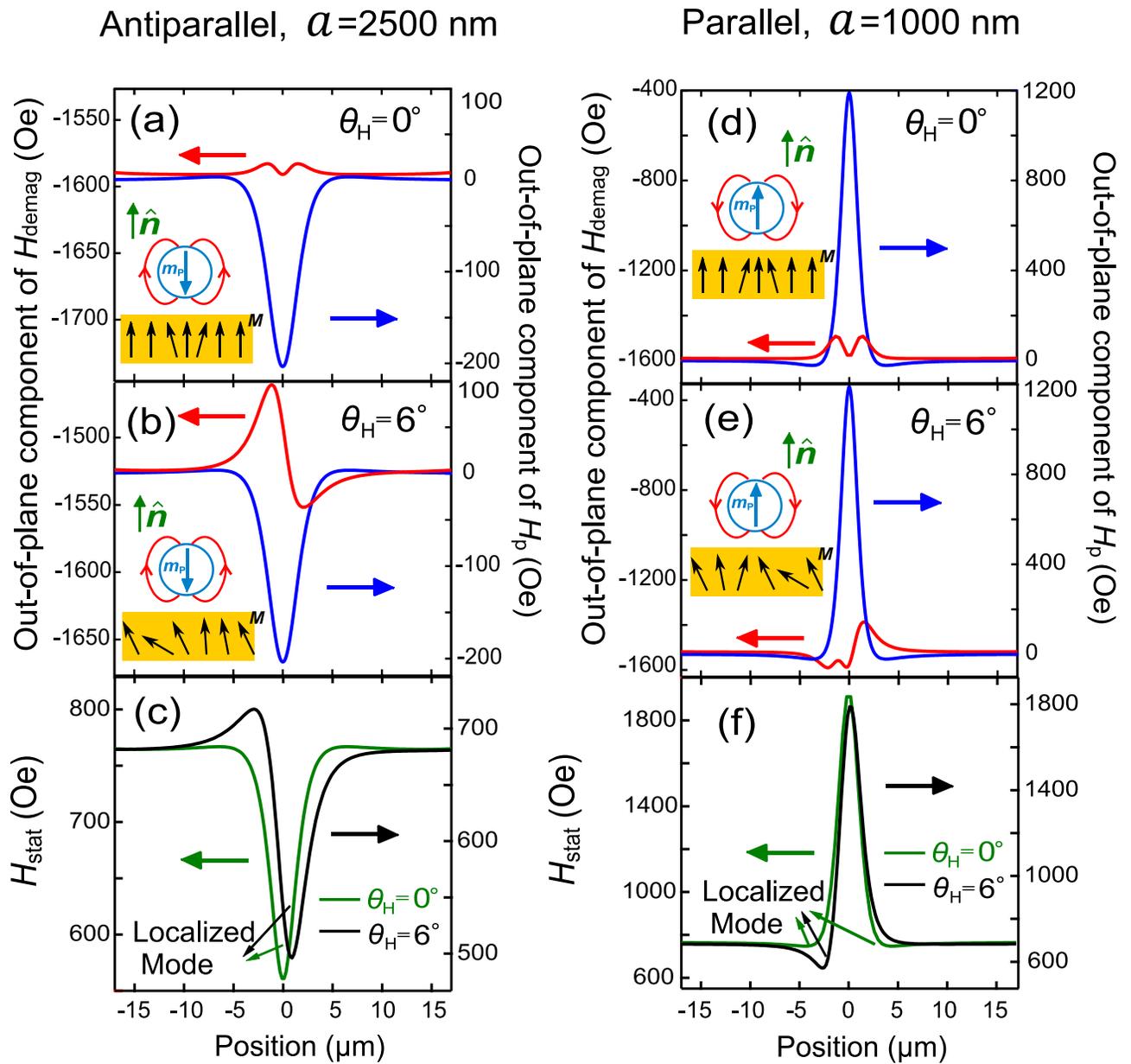

Figure 3.

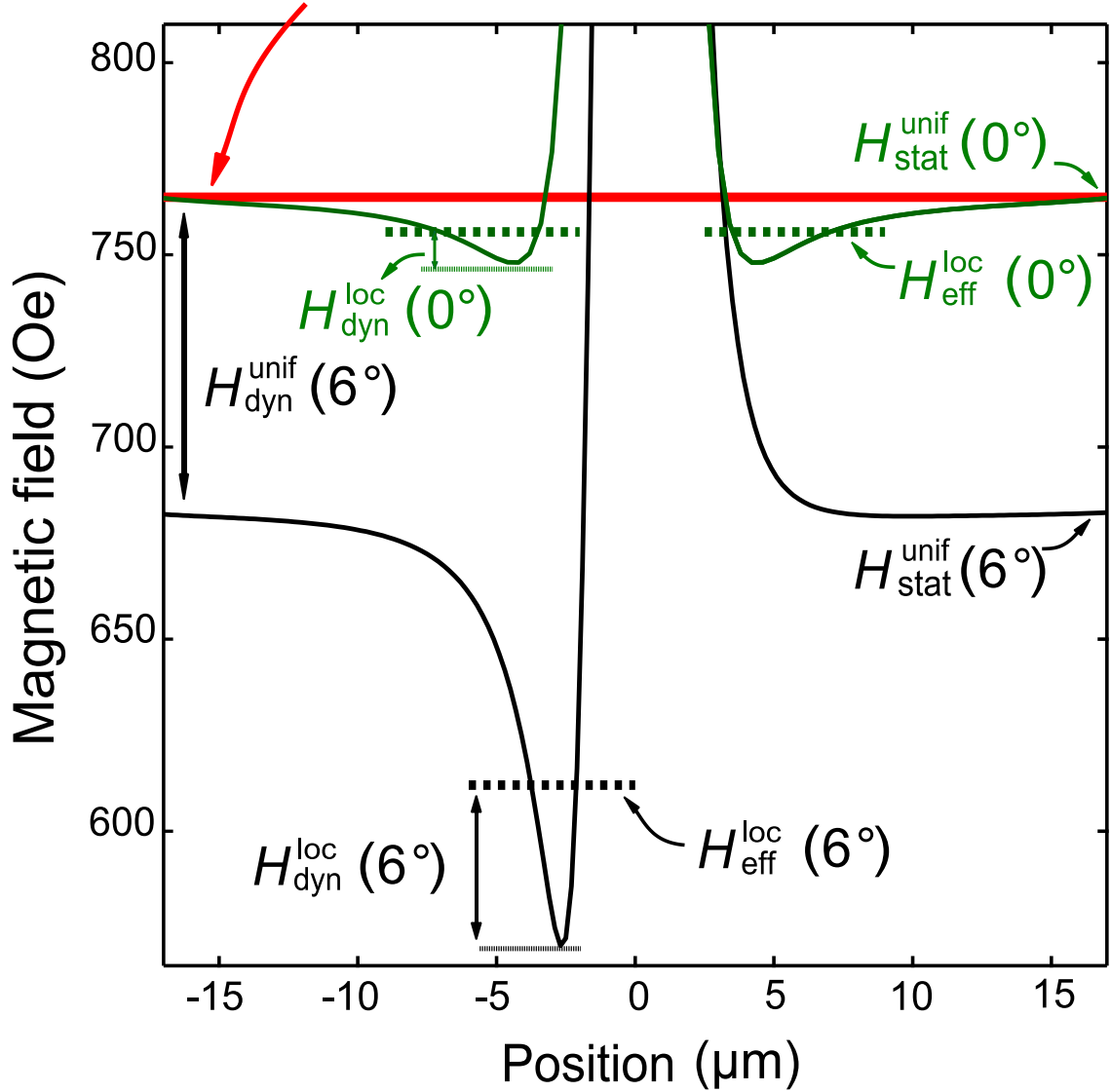

Figure 4.

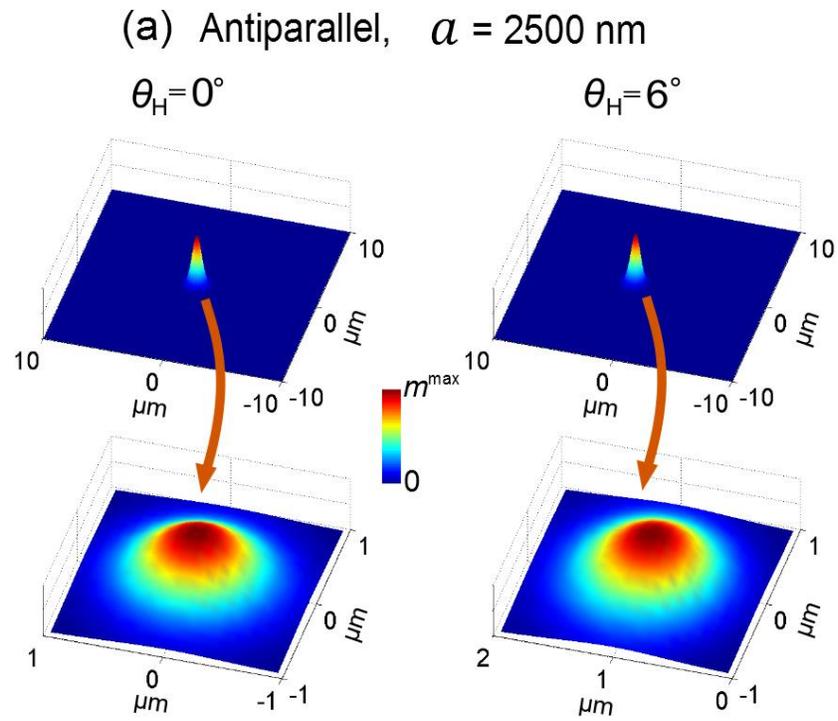

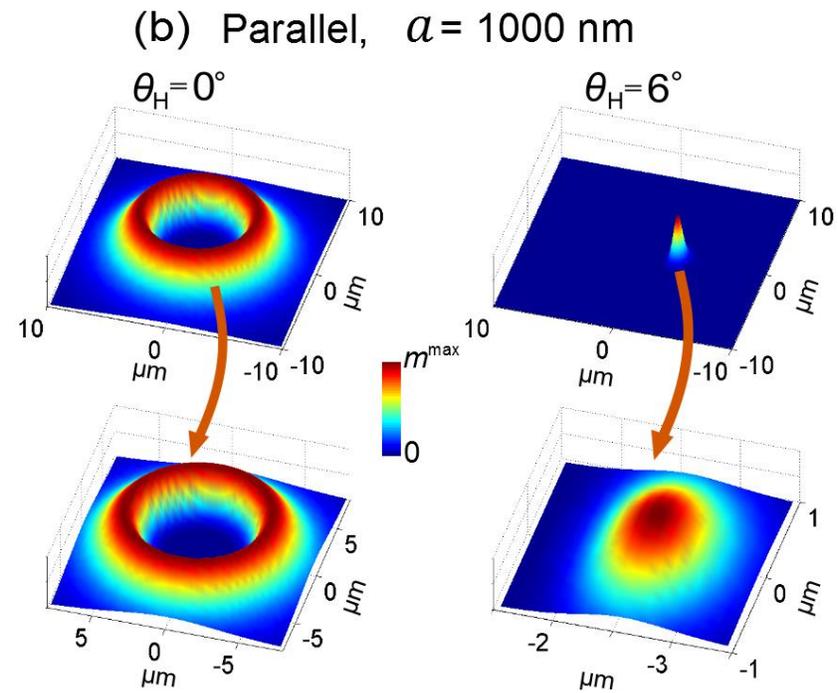

**Figure 5.**

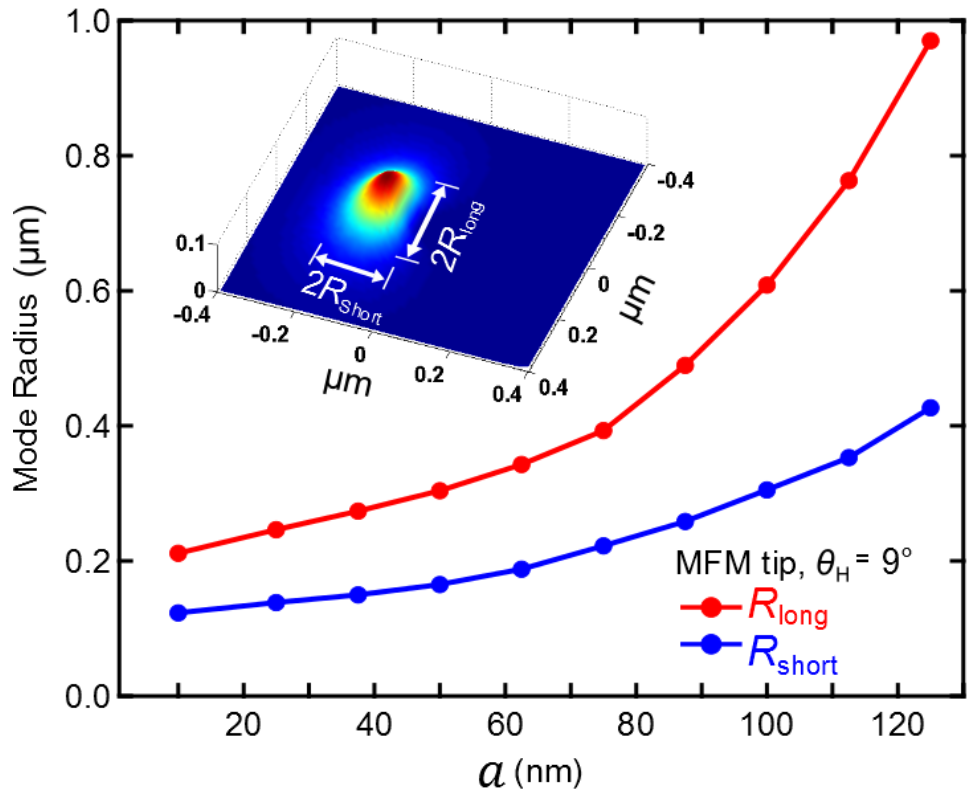

**Figure 6.**